\documentstyle[prl,aps,epsf,epsfig,multicol]{revtex}

\newcommand{\be}{\begin{equation}}
\newcommand{\ee}{\end{equation}}
\newcommand{\bes}{\begin{eqnarray}}
\newcommand{\ees}{\end{eqnarray}}
\newcommand{\bma}{\left( \begin {array}}
\newcommand{\ema}{\end {array} \right)}

\begin{document}

\begin{multicols}{2}

\noindent
{\large \bf Comment on ``Intermittent Synchronization in a Pair of Coupled Chaotic Pendula"}

\smallskip

In \cite{baker}, a number of supposedly novel and surprising features were observed 
in a system composed of two periodically driven and asymmetrically coupled pendula.
In particular it was claimed that `permanent synchronization ... does not occur 
except as a numerical artifact'. It was suggested that this might be related to the 
particular type of coupling. In this comment I want to point out that some of these claims 
cannot be maintained.  The synchronization in this system is precisely of standard blow-out 
type \cite{fuji,fuji2}. The observed intermittency is exactly the on-off intermittency 
well known from the synchronization of multifractal chaotic attractors
\cite{fuji2,piko,heagy}.

The system studied in [1] is described by $\ddot{\theta}_m+\dot{\theta}_m/Q+\sin \theta_m
=\Gamma \cos(\Omega t),\;\ddot{\theta}_s+\dot{\theta}_s/Q+\sin \theta_s
=\Gamma \cos(\Omega t)+c[\sin \theta_s-\sin \theta_m].$ The subscripts $m$ and $s$
stand for master and slave. When they are nearly synchronous, the 
difference $\delta \equiv \theta_m-\theta_s$ satisfies the linearized equation
$\ddot{\delta}+\dot{\delta}/Q+(1-c)\,\delta\cos \theta = 0$. My first observation is that 
the same linearized equation would follw from a symmetric coupling, where master 
and slave have coupling terms $\pm c/2[\sin \theta_m-\sin \theta_s].$ Since the 
behavior near the synchronization threshold is governed by the linearized equation, 
it follows that any eventual abnormal behavior cannot result from the asymmetry of 
the coupling. 

As shown in \cite{fuji}, the synchronization threshold is given by the condition 
that the largest Lyapunov exponent $\lambda_1(c)$ of the linearized equation is zero. In the 
present case this gives $c_c=0.7948$ for the parameter values considered in 
[1]. The eigenvalues of the instantaneous systems 
given in eq.(9) of [1] are irrelevant, except that their fluctuations suggest that 
also the pointwise Lyapunov exponents might fluctuate.
This is indeed the case. Let us 
consider a finite but large time $T$ and define by $\Lambda_1$ and $\Lambda_2$ 
the multipliers along the stable resp. unstable manifold of the linearized equation.
Of course they depend parametrically on the trajectory $\theta(t)$. At the 
synchronization threshold, we have $\langle \log |\Lambda_1|\rangle \equiv T\lambda_1(c)=0$, where 
the average is taken over all initial conditions $\theta(t_0),\dot{\theta}(t_0),
\delta(t_0),\dot{\delta}(t_0)$ with $t_0\ll 0$.
Generically we expect that $\langle \log |\Lambda_2|\rangle /T=\lambda_2(c) <\lambda_1$, 
as is verified numerically. Therefore we have only one direction in the space 
spanned by $(\delta,\dot{\delta})$ along which we must study a possible break up of 
synchronization. 

This break up can occur, even for $c>c_c$, if $\Lambda_1$ 
fluctuates and if the system is perturbed by noise \cite{fuji2,piko}.
More precisely, if this noise is infinitesimal and the fluctuations of $\Lambda_1$
follow normal central limit behavior, one expects intermittent bursts with power behaved 
distributions of amplitudes \cite{piko} and of legths of the locked phase \cite{heagy}.

To describe the fluctuations of $\Lambda_1$ we use the generating function
\cite{piko} $g(z;c)=T^{-1} \log \langle |\Lambda_1|^z\rangle$. The 
cumulant expansion of $\log |\Lambda_1|$ corresponds to a Taylor expansion 
$g(z;c) = z \lambda_1(c) + z^2\sigma^2(c)/2 + \ldots$, where $T\sigma^2(c)$ 
is the variance of $\Lambda_1$, and contributions of higher order cumulants are 
straightforward to compute. The arguments of \cite{piko} can now be used straightforwardly 
to show that amplitudes $\Delta=|\delta|$ of the bursts are distributed according to 
$P(\Delta) \sim \Delta^{-\kappa-1}$ with $g(z=\kappa;c)=0$. Neglecting higher order 
cumulants this gives $\kappa = 2\lambda_1(c)/\sigma^2(c)$. For the 
parameters used in [1], simulations with $T=200$ give 
$\sigma^2(c) = 0.89$ at $c=c_c$, while $\lambda_1(c)\approx 
6.1(c_c-c)$. The predicted power laws for $P(\Delta)$ are compared to numerical simulations 
in fig.1. In the same Gaussian approximation, the distribution for the locking 
intervals $\tau$ is for $c\le c_c$ given by the distribution of return times to a 
reflecting wall of a biased 1-d random walk with drift $\lambda_1(c)$ and diffusion 
constant $\sigma^2(c)$, 
$P(\tau) \sim \tau^{-3/2} \exp(-\tau\lambda_1(c)^2/2\sigma^2(c))$ \cite{feller,heagy}. This 
disagrees with the fit in fig.2 of [1] by the prefactor $\tau^{-3/2}$ which gives indeed 
most of the $\tau$-dependence seen in that figure. 

\begin{figure}[b]
  \begin{center}
    \psfig{file=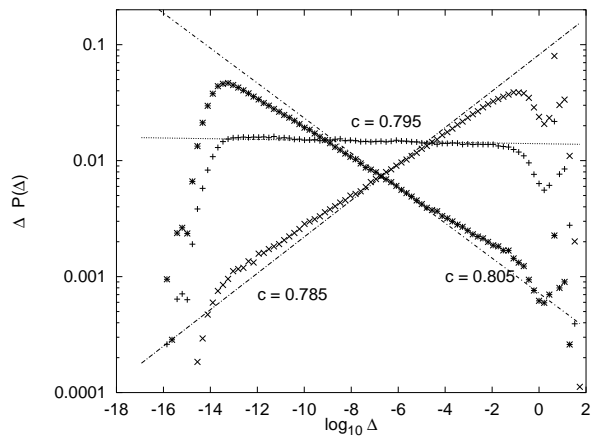,width=6.cm,angle=270}
    \begin{minipage}{8.5cm}
      \caption{Log-log plot of $\Delta P(\Delta)$ for $c=0.795$ (threshold), 
       $c=0.785$ (desynchronized), and $c=0.805$ (synchronized). The system was disturbed by 
       noise with level $10^{-14}$. The straight lines 
       have the theoretically predicted slopes. 
        }
  \end{minipage}
\end{center}
\label{fig1}
\end{figure}

\vspace{-.2cm}

\noindent
Peter Grassberger\\
HLRZ, c/o Forschungszentrum J\"ulich  \\
D-52425 J\"ulich, Germany

Received \today

PACS number: 05.45.+b

\end{multicols}

\end{document}